# Joint Tracking of Groups of Users with Uplink Reference Signals

Henrik Lundqvist, George P. Koudouridis, Xavier Gelabert
Huawei Technologies Sweden AB
Kista, Sweden
{henrik.lundqvist, george.koudouridis, xavier.gelabert}@huawei.com

*Abstract*—In cellular networks user equipment (UE) need to be tracked so that they can be reached by incoming data and to keep context information such as encryption keys available for UE originated transmission. Typically UEs measure reference signal transmissions that are broadcasted by the network, and report to the network based on some criterion that allows the network to know the UE location with sufficient accuracy. An alternative approach is to let the UEs send out reference signals that the network can detect to track the UE location. This reduces the need for the UEs to measure and report, while it requires some resources for uplink transmission. In this paper we propose and evaluate a solution for jointly tracking groups of UEs that are moving together. The results show that UEs can be tracked efficiently with low resource consumption.

## I. Introduction

Mobility is a key feature for radio access networks, and an important differentiator between wide area cellular networks and other radio access network. In the design of the next generation of mobile networks the mobility requirements are getting more stringent, both in terms of reduced interruption time, robustness against failure and energy efficiency. Much of the fragility of the current mobility solution in LTE comes from the signaling of measurement reports and handover command over the air interface at the cell edges [1]. A mobility solution that reduces these transmissions therefore has substantial potential for improving robustness. This can be achieved by letting the network measure uplink (UL) reference signals (RS) instead of relying on the UE measuring and reporting downlink (DL) RS, as in traditional cellular mobility [1]. This may also reduce the energy consumption of the UEs since the UL transmission can give a very short on time during a duty cycle. Since UL channel measurements are needed for massive MIMO it is also beneficial to use the same measurements for mobility purposes. Another benefit of using UL measurements for UE mobility is that it is possible to improve the network performance by network side upgrades without UE impact.

Often UEs are moving together, this has been identified as an enabler for more efficient mobility solutions. In 3GPP a study on mobile relay was conducted to determine the feasibility of providing connectivity to passengers on trains [3]. Large number of correlated handovers can be problematic since it creates high volumes of simultaneous signaling. Also handover strategies taking into account user grouping without the need to deploy mobile relays have been proposed, for example with the goal to distribute the load reasonably [4]. One aspect to be solved is the identification of the users that are moving together, for example passengers on the same train. This could be addressed by analysis of signaling traffic to the core network which does not require access to the radio interface signaling [5]. However, to make use of the user grouping for radio resource management (RRM) purposes the grouping of the UEs should be made in the access network, for example using methods to correlate time and locality information, as described in [6]. Using UL RS allows the network to estimate the position of the UEs without explicitly requesting location information from the UEs, which enables such correlation. Since the UL mobility operation is quite lightweight the group mobility procedures can be applied even for small groups, for example in a car, where both the car itself and devices carried by passengers are connected to the network. By handling the handover of UEs as a group, it is possible to reduce the signaling and to increase the robustness since the process will be similar for all of the UEs in the group. In case the mobility decisions are based on UL RS measurements the resources for UL RS transmission is a critical factor where savings can be made by grouping UEs together and using a single UL RS for the group. In this paper we describe how UEs can be configured with UL RS transmission from one UE which is received both by the other UEs in the group to monitor group membership and by the network to track the group. We also use simulations to quantify benefits in terms of resource consumption and detection performance. Moreover, we discuss constraints that need to be fulfilled for the UE reachability to be handled jointly for a group of users.

In the next section, mobility based on UL beacon tracking is described, while the optimized solution for tracking of groups of users is described in Section III. The feasibility of the solution is evaluated in Section IV together with an analysis of the benefits in terms of signaling reduction. Finally the paper is concluded in Section V.

## II. Mobility Based on UL Measurements

Mobility based on UL RS measurements has been described and the feasibility of the concept has been demonstrated for individual users in [1]. The UL RS consists of Zadoff-Chu sequences transmitted in reserved time-frequency resources. The network needs to assign orthogonal resources to each UE, at least within an area that is larger than the reach of the UL RS transmission, to be able to identify each UE. If the network has sufficient resources assigned for the UL RS transmission and use an adequate assignment of the UL RS to the UEs so that they are orthogonal within a larger area, frequent re-assignment of the UL RS for moving UEs can be avoided. Since it is quite straightforward to make an approximate positioning of the UEs based the UL RS transmission, this can be done efficiently by the network without additional involvement of the UE. However, with a given amount of spectral resources assigned for UL RS transmission there is a tradeoff between how often UEs transmit the UL RS in order to be tracked accurately and the reuse distance of the UL RS resources, where short reuse distance causes increased interference.

The resources that are assigned for UL RS can be distributed in the time frequency plane as illustrated in Fig. 1. A part of the system bandwidth are used for UL RS ($B_{Be}$), at each beacon occasion which occurs with a given periodicity. Each Zadoff-Chu sequence is transmitted in a limited bandwidth ($B_{SEQ}$), which determines the number ($N_{BeRB}$) of beacon resource blocks (BeRB) that can be transmitted in each beacon occasion. In the time domain each transmission requires additional cyclic prefix transmission ($T_{SCP}$) and guard time ($T_{SGT}$) in addition to the time for transmission of the sequence itself ($T_{SEQ}$). A number of orthogonal sequences ($M_{SEQ}$) that are generated from ($|q|$) different root sequences and ($\alpha_{SEQ}$) cyclic shifts of each root sequence can be sent in parallel in the same resource block. Increased transmission time $T_{SEQ}$ or sequence bandwidth $B_{SEQ}$ allows an increased length ($N_Z$) of the Zadoff-Chu sequences, and hence more orthogonal sequences $M_{SEQ}$. By selecting these parameters appropriately the resources for UL RS can be dimensioned.

We consider a time division duplex radio interface and an access network architecture with gNBs that control multiple transmission and reception points (TRP). To handle mobility efficiently for UEs with different communication and mobility patterns different mobility states are defined in mobile communication systems. In 5G connected, inactive and idle mode mobility are defined. Mobility based on UL RS measurement can in principle be applied in both connected and inactive states. In these states a gNB in the access network is responsible for handling the mobility of the UE. To decide which gNB should handle the UE, the network needs to collect measurement results from multiple TRPs and determine which has the best radio channel conditions to the UE. Hence, neighbor gNBs need to exchange measurement results with each other regarding UEs that are moving in the coverage border areas. With an increased accuracy of the tracking the UE can be reached by DL paging or other signaling more efficiently, that is the signaling needs to be transmitted in a smaller area by the network. For inactive mode it may be sufficient to track the users with a granularity of a multi-cell area, while for connected UEs the tracking should be within a cell, and preferably even TRP granularity.

### III. GROUP TRACKING BY UL BEACONS

We consider the tracking of users that do not have any ongoing flows of data transmission; therefore it is not necessary to have detailed knowledge about the channel condition between the user and the TRPs in the network. It is sufficient to track the user so that it can be reached by downlink signaling and that uplink data or signaling can be received by a TRP that recognizes the UE. When one of the UEs in the group needs to send data it will use normal access procedures, e.g. the channel between the UE and the gNB it connects to needs to be properly estimated. However, it can still keep the configuration for the group and apply it when the data transmission ends, hence it remains a member of the group. Since the network tracks the UE location, the required UE context for the data transmission will be available in the gNB that receives an UL transmission, which keeps the latency for the transmission low. In the same way, DL signaling or data is transmitted directly to a UE within a group with identification and resources specific to the UE, and independent of the group. Hence, the grouping of the UEs is only used for efficient location tracking of UEs that are moving together.

The network manages the group by first identifying that UEs are moving as a group. For UEs configured to transmit uplink RS it is relatively simple for the network to estimate the positions and from that determine that a number of UEs behave as a cluster. The network keeps context information that identifies which users belong to a group, and forwards this from gNB to gNB as the group of UEs move in the network. In Fig. 2 the grouping procedure is illustrated, where each UE is transmitting individual UL RS at the beginning of the procedure. The gNB identifies the correlation of the location of the three UEs during a period to determine that the three UEs can be clustered into a group.

Once the network has identified that UEs can be classified as a group, it configures one of the UEs to send group UL RS, in Fig. 2 it is UE2. The other UEs in the group are configured to receive those group UL RS and stop transmitting their allocated individual UL RS. The configurations of the UL RS contain the BeRB, frame number offset, and the Zadoff-Chu sequence that describe the UL RS. The UEs in the group that shall receive the UL RS need to switch to receiving mode while the UL RS of the group is being transmitted. Hence, the configurations of transmission and reception times for the RS are individual for the UEs in the group. Since the UEs need some time to reconfigure between receive and transmit, the network needs to take this into consideration when it schedules DL transmissions or UL resources to the UEs. This is relatively straightforward since it is the network that configures the RS resources for the group. However, the resource configuration needs to be forwarded in the context when the control of the group is handed over from one gNB to another. This way the handover can be hidden from the UEs.

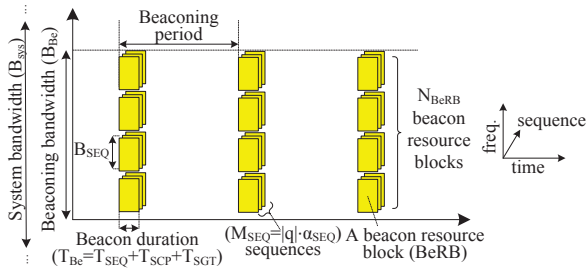

Fig. 1. Beacon resource dimensioning [2].

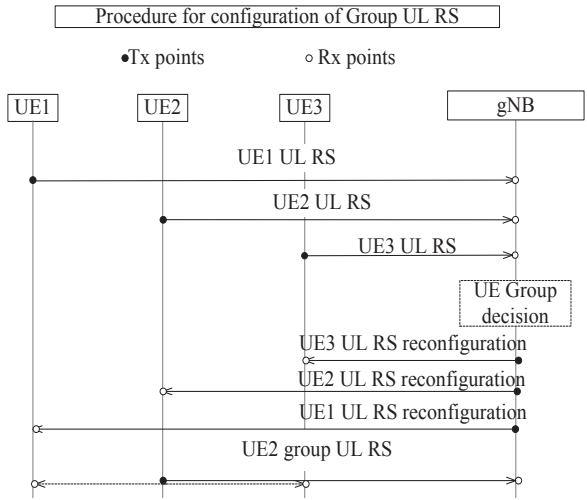

Fig. 2 Configuration of UEs with a group UL RS.

Adding UEs to the group can be made in an analogous way, by first identifying that the UL RS of the group and of the UE are moving in a correlated way. Then the UE can be configured to receive the group UL RS and stop transmitting its own UL RS.

By receiving the group UL RS, the other UEs in the group, UE1 and UE3, can determine that they are still co-located with the group. If a UE can no longer receive the group UL RS, it signals to the network that it is no longer moving together with group. The network then changes the context information to indicate that the UE is no longer part of the group and it configures the UE with new UL RS resources so that it can be tracked individually by its own UL RS transmissions. If it is the UE that is transmitting the group UL RS that leaves the other UEs of the group, all the other UEs could be reconfigured with individual UL RS, or the network may attempt to configure them into a new group. However, since the network cannot reliably determine that they are still moving as a group the safest option is to configure all the UEs with individual beacons and observe the correlation between the received UL RS before it determines if it can still be considered as a group.

Which UE that transmits the UL RS for the group may be reconfigured periodically to distribute the battery consumption for the UL RS transmission fairly among the UEs.

## IV. EVALUATION

The feasibility of the group UL RS solution is evaluated numerically by means of simulations. By grouping the users together, the number of UL RS transmission per UE can be reduced, or alternatively the UL RS can be transmitted at shorter intervals. For the numerical evaluation we assume that the UL RS are transmitted at shorter intervals when the UEs are grouped together.

### A. Simulation model

A user $u_i$ will transmit a beacon using a subset of subcarriers $n_i$ forming a BeRB. The SINR over the subset of subcarriers $n_i$ is given by

$$\bar{\gamma}_{i,j}^{(n_i)} = \frac{\sum_{n \in n_i} P_{b,i}^{(n)} \cdot G_{i,j} \cdot |h_{i,j}^{(n)}|^2}{P_N + \sum_{k \neq i} \rho_{q_i,q_k} \left( \sum_{n \in n_i} P_{b,k}^{(n)} \cdot G_{k,j} \cdot |h_{k,j}^{(n)}|^2 \right)} \quad (1)$$

$$\triangleq \frac{S_{i,j}^{(n_i)}}{P_N + I_{i,j}^{(n_i)}}.$$

where $h_{i,j}^{(n)}$ is the channel gain due to fast fading for the $n$-th subcarrier between user $i$ and AN $j$, $P_{b,i}^{(n)}$ is the beacon power for a user $i$ over subcarrier $n$, $P_N$ is the noise power over the set of subcarriers $n_i$, and $G_{i,j}$ is the channel gain between user $i$ and AN $j$ obeying log-normal shadow fading. The interference $I_{i,j}^{(n_i)}$ from other users' sequences depends on the cross-correlation parameter $\rho_{q_i,q_k}$ given by

$$\rho_{q_i,q_k} = \begin{cases} \delta(c_i - c_k) & , r_i = r_k \\ \frac{1}{\sqrt{N_Z}} & , r_i \neq r_k \end{cases} \quad (2)$$

where $\rho_{q_i,q_k}$ is a correlation factor between two sequences $q_i$ and $q_k$ sent by users $u_i$ and $u_k$, $(c_i - c_k)$ denotes the cyclic shift difference between sequences $q_i$ and $q_k$, and $\delta(x)$ is the delta function.

Assuming a flat fading channel, the miss-detection probability can be expressed as a function of SNR as follows [7]:

$$P_{MD} = \left[ F_{\chi^2}(\lambda, 2N_a, 0) \right]^{D-1} \times F_{\chi^2}\left( \frac{\lambda}{1 + N_Z \bar{\gamma}_{i,j}^{(n_i)}}, 2N_a, 0 \right). \quad (3)$$

where $F_{\chi^2}\left( \frac{z}{\sigma^2}, r, \frac{s^2}{\sigma^2} \right)$ is the cumulative density function (CDF) for a (non-) central chi-squared distributed random variable with number of degrees of freedom $r$, non-centrality parameter $s^2$ and where all terms in this chi squared random variable have identical variances $\sigma^2$. $N_a$ is the number of receiving antennas (SISO implies $N_a = 1$), $D$ is the length of the detection search window, and $\lambda$ is the detection threshold. Furthermore, to enforce that the sequence design applies for a certain beacon range, it is assumed that beyond this range the probability of detection is zero i.e., $P_{MD} = 1$.

For the evaluation we consider a system operating at 3.5 GHz with a system bandwidth $B_{SYS}$ of 100 MHz. The number of BeRBs is 53 and the total number of sequences 636. The maximum beacon rate per user and second is set to 5, corresponding to a beacon interval of 200 ms. The time-domain and frequency-domain signature sequence parameters and other

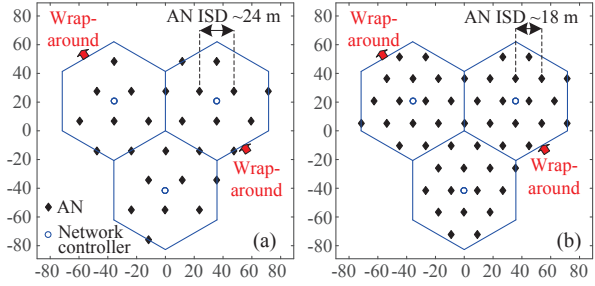

Fig. 5 Scenario deployment for beacon miss-detection and false-alarm cases.

TABLE I. PERFORMANCE EVALUATION PARAMETERS.

| Parameter | Value |
|---|---|
| Center frequency | 3.5 GHz |
| System BW ($B_{sys}$) | 100 MHz |
| Beacon resource blocks ($N_{BeRB}$) | 53 |
| Beacon duration ($T_{Be}$, $N_{Be}$) | 4.74 μs, 1164 samples |
| Sequence duration ($T_{SEQ}$, $N_{SEQ}$) | 4.17 μs, 1024 samples |
| Sequence subc. Spacing ($\Delta f_{SSC}$) | 240 kHz, |
| Sequence CP duration ($T_{SCP}$, $N_{SCP}$) | 0.35 μs, 86 samples |
| Sequence GT duration ($T_{SGT}$, $N_{SGT}$) | 0.22 μs, 54 samples |
| Beacon range ($R_{F_i}$ $i = 0,1,2,3$) | 32.9 m |
| Sequence length ($N_Z$) | 7 samples |
| Number of root seq. ($|q|$) | 6 |
| Cyclic shifts per root seq. ($\alpha_{SEQ}$) | 2 |
| Total seqs. per BeRB ($M_{SEQ}$) | 12 |
| Total seqs. in $B_{sys}$ ($M_{sys}$) | 636 |
| UL Beacon power ($P_{b,i}$) | 15 dBm |
| `ReuseDistThreshold` | 20 m |
| Thermal noise over BeRB | $P_N = kTNFB_{SEQ} = -72.74$ dBm<br>$k = 1.38 \cdot 10^{-23}$, $T = 290$,<br>$NF = 9$dB |
| Channel model: 3D UMi [8], with log-normal shadow fading $\sigma_{SF} = 3$ dB (LoS) and $\sigma_{SF} = 4$ dB (NLoS). Correlated Rayleigh fast fading mean=1 | |

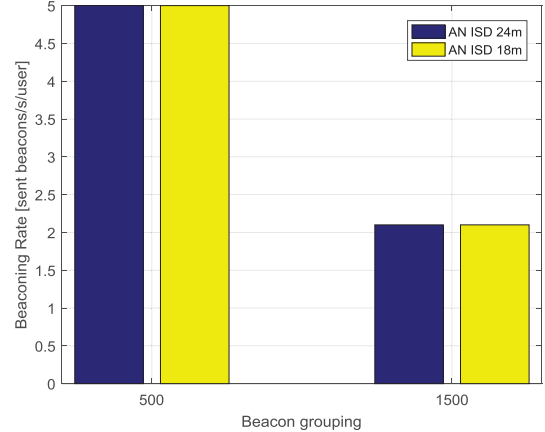

Fig. 3 Beaconing rate against number of users for different ISDs

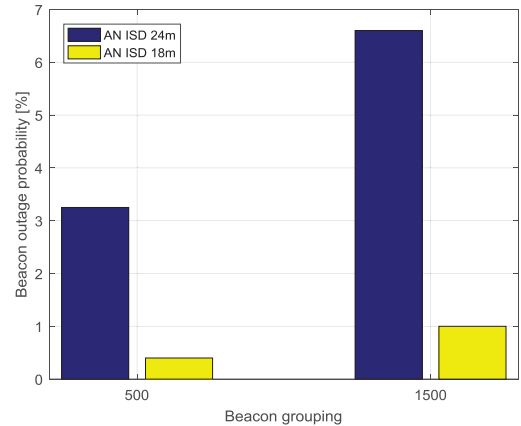

Fig. 4 Beacon miss-detection probability against number of users for different ISDs.

propagation model parameters used in the simulations are summarized in TABLE I.

The simulated network deployment is illustrated in Fig. 5. Users, who transmit their beacons at 15 dBm transmit power, move in a random directions at a speed of 30 km/h. Wrap-around is used to remove border effects related to interference and user movement. The total number of users is 1500. Two different deployments with two different inter-site-distances (ISDs) of roughly 24m and 18m respectively are simulated. For each deployment two different beaconing schemes are compared. The first beaconing scheme corresponds to a beacon grouping scheme where the 1500 users forms 500 groups of three users each. At any time one user from within the group transmits group UL beacons representing all users in the group. This is a typical urban dense scenario where users are situated in vehicles driving in a street of a busy area and where the velocity of the vehicles do not exceed 30km/h on average. This scheme is compared to a regular beaconing scheme where all users are transmitting UL beacons individually.

### B. UL transmission performance

Fig. 3 shows the beaconing rate per user against the user load for different deployments. For the 500 groups, the available beaconing resources allow for the maximum beacon rate for both network deployment scenarios. For the 1500 individual users, although spatial reuse provides beaconing rate gains (with respect to no spatial reuse) of around 27%, the beacon rate decreases significantly. Alternatively, this would require the utilisation of approximately 2.5 times more beacon resources to track all 1500 users as compared to the 500 groups of users if the beacon interval should be kept the same.

Fig. 4 plots the miss-detection rate, i.e. the probability that a beacon transmission is not detected by any single access node. It can be seen that the miss-detection rate is reduced by roughly

50 percent when the UEs are grouped. This is an effect of the reduced interference due to the reduced UL RS transmission.

*C. Implications for UE reachability*

When data destined to a UE arrives to the mobile network, the network needs to signal this to the UE in the DL over a paging channel or another control channel that the UE listens to regularly. The more accurate the UE location information is, the more resource efficient the DL signaling can be since the transmission can be constrained to a smaller area. For the group tracking to work, all the UEs in the group need to be able to receive the same DL control channel. Hence, the group cannot extend too much spatially. Typically paging for inactive UEs may be sent over multiple cells, while DL control information for connected UEs will be sent in a single cell and could even be limited to a subset of TRPs or beams in the cell if the network is tracking the UE location with sufficient accuracy. In the following we assume that the DL control information is transmitted in one cell.

It is non-trivial to determine the actual reachability over the DL control channel since the interference situation is not the same for the UL and the DL channels. The interference in the DL depends on the DL control channel design which is out of scope for this paper. However, to give an idea of the feasibility of the UE group tracking we evaluate the probability that a UE in a group would have lower path loss to a different cell than the UE that transmits the UL RS. Fig. 6 shows this probability as a function. It can be seen that already with a group radius of 5 meter around 20 % of the UEs may have better connection to a different cell. However, this is not a problem as long as the network can reach the UE from the cell it selects for the DL transmissions based on the group UL RS. For a group radius of 5 meters the average difference in path loss for UEs in the group compared with the group UL RS is around 3.5 dB, which would typically leave good margins in such a dense network. In a less dense network the probability of connecting to the wrong cell would be reduced since the cells are larger.

It is also worth noting that the channel difference within the group is not the only inaccuracy that may cause the network to use the wrong cell to reach the UE. For example, for a UE moving at 10 m/s the inaccuracy of the UE tracking would be up to two meters due to the mobility assuming a UL RS interval of 200 ms. If the UEs would use individual UL RS rather than group UL RS it would be necessary to extend the interval due to the limited resources, hence the inaccuracy due to mobility would increase when the inaccuracy due to grouping is eliminated.

V. CONCLUSION

Mobility based on UL measurements has good potential to handle the mobility challenges occurring with the new technology in 5G. A challenge for UL measurement based mobility is the availability of resources for transmission of UL RS. Therefore, it is beneficial to group the UEs into clusters that are moving together and track each group by a single UL RS. This paper describes how this can be done by configuring one UE in the group to transmit the UL RS and the others to identify that they are co-located with the transmitting UE by receiving the UL RS. The reduction of the required UL RS resources reduces the interference and allows the RS to be transmitted more frequently.

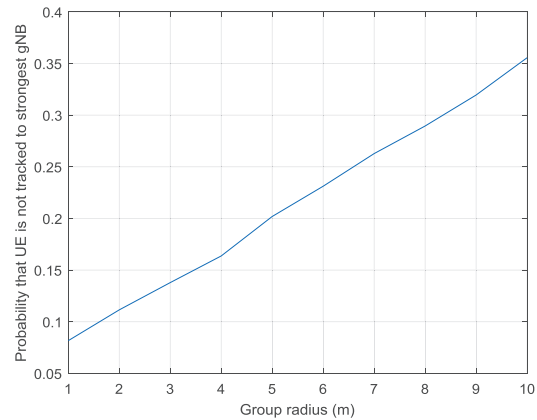

Fig. 6 The probability that a UE in the group has lower path loss to a different cell than the beaconing UE, as a function of the maximum of a group radius centered at the UE transmitting the UL RS.

The solution will be further evaluated for less dense network deployments, and the performance in terms of paging miss rate compared to both traditional DL measurement based mobility and UL based mobility without UE grouping.

VI. ACKNOWLEDGEMENT

This work has been performed under grant agreement H2020-MSCA-ITN-2016 SECRET-722424 funded by the European Union's Horizon 2020 research and innovation programme.